\documentstyle[twocolumn,aps,prl]{revtex}

\begin{document}

\twocolumn[\hsize\textwidth\columnwidth\hsize\csname %
@twocolumnfalse\endcsname
\draft
\title{Hidden Dimer in the Frenkel-Kontorova Model}
\author{ Jukka A. Ketoja}
\address{Department of Mathematics, University of Helsinki, P. O. Box 4,
FIN-00014 Helsinki, Finland}
\author{Indubala I. Satija\cite{email}}
\address{
 Department of Physics, George Mason University,
 Fairfax, VA 22030}
\date{\today}
\maketitle
\begin{abstract}
The incommensurability of the supercritical Frenkel-Kontorova model
is decomposed into a family of $dimer$ type ``defects'' by appropriate
decimations. Interestingly, this hidden dimer results in Bloch-wave type 
excitations of the renormalized chain   
which appear in the disguised form of multi-step
phonon modes for the original system. We call these intriguing
excitations {\it stepon modes} and conjecture that they play a key role
in determining the localization boundary in systems exhibiting Anderson
localization. 

\end{abstract}

]

\narrowtext

In this paper, we describe a very interesting relationship
between two seemingly different
models that have been in the forefront of theoretical physics:
the Frenkel-Kontorova (FK) model \cite{Aubry} and
the random dimer model \cite{dimer}.
This relationship establishes the existence of propagating Bloch-type phonons
in the pinned phase of the FK model and resolves the mystery of
novel multi-step excitations found recently \cite{KSphon}.

The incommensurate FK model consists of a chain of balls connected
by Hooke's springs in a periodic sinusoidal potential of the strength $K$
where the average spacing $\sigma$ between the balls is incommensurate
with the periodicity of the potential. 
The equilibrium positions of the balls $x_n$ are
given by the iterates of the standard map 
\cite{Aubry}
\begin{equation}
x_{n+1}+x_{n-1}-2x_n = - {K\over  2\pi } \sin( 2 \pi x_n) .\label{SM}
\end{equation}
For small $K$, the iterates $x_n$ are confined on an invariant circle
of the standard map. The breakup of the invariant circle
at the critical value $K_c$
corresponds to the pinning transition in the FK model accompanied
with the disappearence of the zero frequency phonon mode.
The phonon modes are linear excitations of the equilibrium configurations
of the balls satisfying the equation
\begin{equation}
\psi_{n+1} + \psi_{n-1} + \epsilon_n \psi_n = E \psi_n , \label{phonon}
\end{equation}
where the eigenvalue $E$ is related to the phonon frequency $\omega$ as
$E=-\omega^2+2$, and
$\epsilon_n = K\cos(2\pi x_n )$ is the onsite phonon potential.

At the so-called anti-integrable limit $K \rightarrow \infty $ \cite{AA},
the balls sit at the bottom of the potential wells and therefore
$\epsilon_n$ is constant for all sites $n$. When $K$ decreases,
the limiting uniform phonon potential splits into more and more 
distinguishable levels. As shown in Fig. 1, this splitting 
generates a symbolic representation of the lattice. 
It turns out that by expanding the potential upto any order $1/K^p$
the resulting incommensurate lattice with no apparent short range clustering
can be decimated into a lattice with dimer and tetramer ``defects''.
As $p$ increases, the decimated structure becomes more and more complex.
The discovery of a hidden dimer helps in tracing the
origin of intriguing ``stepon'' excitations with multi-step
eigenfunctions that were found
in our recent numerical study of the supercritical parameter region
$K > K_c$ \cite{KSphon}.
The central result of this paper is that the stepon modes of the FK model
are in fact propagating Bloch waves on the renormalized lattice.

The models with dimer type defects attracted a great deal of attention a
few years ago \cite{dimer} due to the discovery of subtle
delocalization mechanisms which explained the conduction in polymers. 
The delocalized modes originated from special
resonance conditions where the reflected waves from neighboring defect sites
interfered destructively resulting in the ballistic transport in the system.
Detailed theoretical studies showed that a variety of extended defects with
reflection symmetry could be made to exhibit this resonance by tuning a control
parameter so that the resonance condition was satisfied by an energy close
to the Fermi energy. This turns out to be true also for
the dimer and tetramer defects in the renormalized lattices of
the supercritical FK model. 

For our analysis, we study a generalized version
of the FK model with two independent parameters $K$ and $\lambda$ so that
in Eq. (\ref{phonon}) $\epsilon_n =\lambda \cos(2\pi x_n )$.
The model reduces to the FK model when the two parameters are equal.
This extension of parameter space is analogous
to the analytic continuation into complex plane and will turn out to be crucial
in understanding the stepon modes.
In the 2-parameter model, the parameter $\lambda$ controls
the strength while $K$ determines
the smoothness of the onsite ``phonon'' potential as the underlying
invariant circle of the standard map
undergoes the transition by breaking of analyticity 
at the critical value $K =K_c$ \cite{TBA}. In the region $K < K_c$, the 
potential $\epsilon_n$ is a smooth function of the effective phase
$\theta = \{ n\sigma + \phi \}$ where the brackets denote the fractional part.
In this case the model falls into the universality
class of the famous Harper equation \cite{Harper}
which exhibits a transition from extended to localized states.
On the other hand,
for $K > K_c$ the phonon potential is determined by an
underlying invariant Cantor set (cantorus) of the standard map.
Therefore, the 2-parameter model provides one with a new class
of quasiperiodic systems that interpolates nicely between the
Harper equation ($K=0$) and the FK phonon equation.
As seen later, this important feature helps in understanding 
the absence of Anderson localization for the FK phonons.

Fig. 2 shows the phase diagram of the model
obtained using an exact decimation
scheme discussed in our earlier papers \cite{KSphon,KSloc}.
The model exhibits an extended phase, a localized phase
as well as a critical phase with self-similar eigenfunctions.
A novel aspect of the model
is the emergence of infinitely many curves from the ``corner''
$\lambda,K \to \infty$ along which the eigenfunctions 
are represented by an infinite series of step functions of the effective
phase $\theta$ (see Fig. 3). We call these solutions 
$stepon$ modes. All except the right most curve intersect the standard FK limit
$K=\lambda$. We referred to these intersections as degeneracy
points in our earlier numerical work \cite{KSphon}
because at these parameter values, the nontrivial scaling properties of 
critical phonons ``degenerated'' into the trivial scaling of
the stepon modes. It is interesting to note that
the degeneracy curves accumulate at the
critical parameter value $K=K_c$ for the circle-cantorus
transition of the standard map. The parameter region outside these curves 
consists of either the critical or localized phase. The localized
phase for large values of $\lambda$ intertwines with the 
degeneracy curves. The fact that the localized modes 
reside inside the $tongues$ of the degeneracy curves
suggests that the stepon modes play a special role
in the localization phenomenon in models where the underlying
potential is not smooth. However, the specifics of that role
remains eluded to us at present.

The above phase diagram suggests a perturbative approach
to understand the origin of the stepon modes.
We develop a perturbation theory near the anti-integrable limit
by expanding the equilibrium configurations
of the balls $x_n$ with $\kappa = 1/K$ as the expansion parameter.
In the cantorus regime, the hull function $X$ defined by 
$x_n =X(n\sigma +\phi)$
is a convergent series of step functions \cite{Aetal}.
We write
$X(\theta )=X_0 (\theta ) + X_1 (\theta) \kappa + X_2 (\theta) \kappa^2 +...$,
where $X_0 (\theta) = {1\over 2} + Int(\theta )$.
Substituting the above expansion into Eq. (\ref{SM}),
we generate a similar expansion for the ``phonon'' potential
$\lambda \cos[2\pi X(\theta )]$. 
Truncating this expansion at the order $\kappa^p$
leads to a potential with a finite number of steps.
This in turn leads to symbolic representation of the lattice
of the type shown in Fig. 1, where the number of required symbols increases
with the order of perturbation.

For general $p$, the symbolic representation
can be very complex. However, it turns out that by an appropriate decimation,
the lattice can always be represented by one of the two possible
forms which happen to be the ones found at $p=2$ and $p=3$.
Different $p$ cases are distinguished
from each other by different renormalized onsite
``energies'' and coupling terms. In the following,
we summarize the essential features of the symbolic representation.
 
Each symbol (except the one for the potential minimum)
is associated with two symmetrically placed $\theta$-intervals where
the potential takes the corresponding constant value.
For example, for $p=2$ the symbol $a$ is attached to the
$\theta$-intervals $[0,\sigma )$, $[1-\sigma ,1)$ and the symbol $b$ to
the interval $[\sigma, 1-\sigma)$. As $p$ increases, these intervals
split into subintervals in a systematic way.
The borders of intervals, which are discontinuities of the
potential, are obtained as the first $p-1$ forward and backward 
iterates of the map $\theta_{j+1} =\theta_j + \sigma$ (mod $1$) with
$\theta_0 =0$. The splitting of the levels as $p$ increases
can be shown to follow
two possible patterns depending upon whether the new discontinuities
land within the middle interval or not. If the two new discontinuities
at the order $p$ land inside order-$(p-1)$ intervals
other than the middle one, both intervals are split into
two parts whose relative lengths are determined by the 
golden mean. We label the longer ones of these new subintervals
by $a$ and the shorter ones by $b$.
The rest of the intervals can be labeled in an arbitrary way.
This pattern emerges for $p=3,4,6,7,...$.
The other possibility is that both new discontinuities land inside the
middle interval which breaks up into
three parts $a$, $b$, and $a$.
In this case the total length of the $a$-subintervals is related
to the length of the $b$-subinterval by the {\it square}
of the golden mean. This happens to be the case for $p=2,5,18,...$.

It turns out that the lattice sites which are not labeled as $a$ or $b$
can be decimated. In the renormalized lattice, 
the complexity of the original symbol dynamics
manifests itself in the renormalization of
the onsite energies as well as the coupling terms.
Because of the symmetric alignment of the decimated blocks
\cite{blocks},
the renormalized couplings between the $a$-type sites alternate between
two values. Thus, the whole problem reduces to showing 
the existence of propagating wave solutions for the lattice
with dimer and tetramer ``defects''.

It suffices to discuss the $p=3$ perturbation theory in detail ($p=2$
gives the trivial solution)
as the other renormalized lattices have a similar form.
The truncated potential in the $p=3$ case
takes three distinct values $\epsilon_a$, $\epsilon_b$, and
$\epsilon_c$ as shown in Fig. 1. Here
$\epsilon_a=\lambda(-1+2\pi^2\kappa^2(1-6\kappa))$,
$\epsilon_b=\lambda(-1+2\pi^2\kappa^2(1-8\kappa))$, 
and $\epsilon_c=-\lambda$. 
The decimation of the sites $c$ results in the renormalized lattice
$...aaaabbaabb...$
with the renormalized onsite energy $\bar \epsilon_a
=(1-(E-\epsilon_a )(E-\epsilon_c ))$,
while the renormalized couplings between two neighboring
$a$-sites alternate between $V_1=1$
and $V_2=(E-\epsilon_c)$.
The decimated lattice can be shown to have a traveling wave
solution provided the parameters satisfy the condition
\begin{equation}
(\epsilon_b-\epsilon_a)(E-\epsilon_a)(E-\epsilon_c)
=(\epsilon_b-\epsilon_a)+(\epsilon_c-\epsilon_a) . \label{condition}
\end{equation}
Each decimated $c$-site gives rise to the phase shift $\Omega$ given by
\begin{equation}
e^{i\Omega} =\frac{(\epsilon_b-\epsilon_a)e^{-ik}+1}
{(\epsilon_b-\epsilon_a)e^{ik}+1} ,\\
\end{equation}
where $E = \epsilon_b + 2cos(k)$.
Assuming that these phase shifts amount to the total phase shift
$\Omega (n)$ at the $n$th site of the renormalized lattice, the
traveling wave solution can be expressed as $e^{i[kn+\Omega (n)]}$.

It should be noted that within the defects,
the form of the wave is rather complicated.
However, by decimating the sites labeled
by $c$, the complexity has been absorbed into the renormalization of the onsite
energies and the coupling terms.
For $k=0$, we always have $\Omega=0$ except when $\epsilon_b-\epsilon_a =-1$
for which the $k=0$ solution corresponds to $\Omega=\pi$. This special 
solution is actually very similar to the one obtained by
Dunlap et al.\cite{dimer} in their study of the random dimer model.
We would also like to 
point out that our general solution is valid for the type
of defects we discuss irrespectively of whether they originate
from quasiperiodic, chaotic or correlated random processes.

The existence of numerically obtained stepon modes of the type described
in Fig. 3 implies that
delocalized modes exist also for the system obtained by taking into account
the full expansion upto infinite order. In this case,
the stepon eigenfunction is represented by an infinite
series of step functions. Therefore, the stepon modes of the FK model
can be thought of as Bloch waves on an infinitely many times
renormalized lattice. 

Another important result of our analysis is the fact that
it predicts the asymptotic form of the degeneracy curves
shown in Fig. 2. Using the explicit values of the onsite energies
results in the following relationship between the
two parameters $\kappa$ and $\lambda$ for the three rightmost branches:
\begin{eqnarray}
\lambda &=& {1\over 4\pi^2 } \kappa^{-3} + {\cal O} (\kappa^{-2} ) \label{s1}\\
\lambda &=& {1\over 4\pi^2 } \kappa^{-4} - {1\over 4\pi^2 } \kappa^{-3}
+ {\cal O} (\kappa^{-2} ) \label{s2} \\
\lambda &=& {1\over 4\pi^2 } \kappa^{-4} + {\cal O} (\kappa^{-2} ) \label{s3}.
\end{eqnarray}
Eq. (\ref{s1}) was obtained  using the $p=3$ perturbation theory
(Eq. (\ref{condition}))
while the $p=4$ theory gives all three solutions (\ref{s1}-\ref{s3}).
We conjecture that the rest of degeneracy curves can be
explained by the higher order perturbation theory.

In summary, our numerical results along with the
decimation and the systematic perturbation theory demonstrate that
the incommensurate FK model in the pinned phase 
is related to the class of models with correlated defects
and exhibits propagating wave solutions on the renormalized lattice.
Moreover, the extended two-parameter model is the first known case
which exhibits both correlated defects and the localization transition.
The propagating stepon modes seem to play a special
role in determining the onset to Anderson localization.
Proper understanding of how the stepon modes determine
the localization boundary may be difficult as the
perturbation theory in any finite order
does not predict the existence of exponentially
localized modes in the model.

The study of the two-parameter model
helps one to understand an important distinction between
the FK model and the Harper equation.
First of all, the absence of localization in the
FK model is due to the fact that
the nonlinear potential is not strong enough to localize the phonons.
The extended FK model, where the strength and the smoothness are 
controlled independently, does exhibit the localization
transition for $\lambda >> K$.
However, the localized phase for $K> K_c$
exhibits an important distinction from that of the Harper equation.
The localized eigenfunctions (once the exponentially decaying part
is factorized out) for $K < K_c$ exhibit universal self-similar
fluctuations characterized by a unique strong coupling
renormalization fixed point \cite{KSloc}.
In contrast, these fractal fluctuations 
in the cantorus regime $K > K_c$ appear irregular 
and defy any simple renormalization explanation.

After the completion of this work, we found other papers where
extended states have been seen in other aperiodic lattices \cite{TM}. 
However, the propagating modes in the FK model
are somewhat unique as the aperiodicity of the FK model
comes from a mismatch of the continuous
sinusoidal potential with the periodicity of the lattice.
The existence of propagating modes in the pinned FK model
suggests the possibility of extended states in
other pinned systems such as
pinned flux and vortex lattices.
Finally, the FK model describes a variety of systems such as
adsorbed monolayers on substrates and
charge density wave conductors \cite{Su,Aubry}, 
and we hope that our study will provide a
new direction of research in complex aperiodic systems.

The research of I.I.S. is supported by National Science
Foundation Grant No. DMR~097535.  
JAK would like to thank for the hospitality
during his visits to the George Mason University and for
the support by the Magnus Ehrnrooth Foundation.

\begin{figure} 
\caption{Schematic diagram showing the equilibrium ball configurations 
in the incommensurate FK model for $K$ large enough so that the effects of
the order ${\cal O} (K^{-(p+1)})$ are negligible.
(a) $p=2$ and the resulting incommensurate phonon potential 
has two distinct levels which leads to the symbolic representation
$...abaabaaaabaaa...$ of the lattice. (b) $p=3$ and there are
three levels which generate the symbol sequence
$...acaacabbacabb...$. By decimating the sites $c$ out, one obtains the
renormalized lattice $...aaaabbaabb...$ with dimer ($aa$) and tetramer
($aaaa$) ``defects''.
} 
\label{fig1}
\end{figure}

\begin{figure}
\caption{Phase diagram in the $K-\lambda$ parameter plane for $E=E_{max}$
and $\sigma=(3-\sqrt 5)/2$. Here $r=2\arctan(K)/\pi$ and
$t=2\arctan(\lambda)/\pi$.
The dashed line corresponds to the standard FK limit.
The smooth curves show the parameter values for the existence of stepon modes
with $k=\Omega =0$.
The shaded part represents the phase where the eigenfunctions
are exponentially localized.
In the light shaded regime $(K < K_c )$, the fluctuations are self-similar
and fall into the universality class of the Harper equation whereas
in the dark shaded regime $(K>K_c)$ the fluctuations appear irregular.
The unshaded part for $K < K_c$
corresponds to smooth extended eigenfunctions 
while for $K > K_c$ the unshaded regime describes the critical phase.
}
\label{fig2}
\end{figure}

\begin{figure}
\caption{The stepon solution at the first degeneracy point
$r=t=0.67052674...$. Here $\theta= \{ n\sigma -\sigma/2 \}$.
The stepon mode has its discontinuities at the
same locations as the underlying phonon potential.}
\label{fig3}
\end{figure}


\begin{references}

\bibitem{email} e-mail: isatija@sitar.gmu.edu.

\bibitem{Aubry} S. Aubry, Physica D 7, 240 (1983).

\bibitem{dimer} D. H. Dunlap, H-L. Wu, and P. W. Phillips,
Phys. Rev. Lett. 65, 88 (1990);
P. Phillips and H-L. Wu, Science 252, 1805 (1991);
H-L. Wu, W. Goff, and P. Phillips, Phys. Rev. B 45, 1623 (1992).

\bibitem{KSphon} J. A. Ketoja and I. I. Satija, Physica D 104, 239 (1997).

\bibitem{AA} S. Aubry and G. Abramovici, Physica D 43, 199 (1990).

\bibitem{TBA} S. Aubry and G. Andr\'e, Ann. Israel Phys. Soc. 3, 133
(1980); M. Peyrard and S. Aubry, J. Phys. C 16, 1593 (1983);
S. N. Coppersmith and D. S. Fisher, Phys. Rev. B 28, 2566 (1983);
R. S. MacKay, Physica D 50, 71 (1991).

\bibitem{Harper} P. G. Harper, Proc. Phys. Soc. London A 68, 874 (1955);
B. Simon, Adv. Appl. Math. 3, 463 (1982).


\bibitem{KSloc} J. A. Ketoja and I. I. Satija, Phys. Rev. Lett. 75,
2762 (1995); Physica D 109, 70 (1997).

\bibitem{Aetal} S. Aubry, J.-P. Gosso, G. Abramovici, J.-L. Raimbault,
and P. Quemerais, Physica D 47, 461 (1991).


\bibitem{blocks} For $p=3,4,6,7,...$, the general form of the
symbol sequence is $...aCaAaCaAbBbAaCaAbBbA...$, where
$A$, $B$, and $C$ denote arbitrary (sometimes
empty) blocks of other symbols than $a$ and $b$.
For $p=2,5,18,...$, the symbol sequence has otherwise
the same form as in above except that the $bBb$ block is replaced by 
a single $b$.

\bibitem{TM} A. Chakrabarti, S. N. Karmakar, and R. K. Moitra,
Phys. Rev. B 50, 13276 (1994); Phys. Rev. Lett. 74, 1403 (1995).

\bibitem{Su}
S. Coppersmith, T.C. Jones, L.P. Kadanoff, A. Levine, and J.P. McCarten,
Phys. Rev. Lett. 78, 3983 (1997).


\end{references}
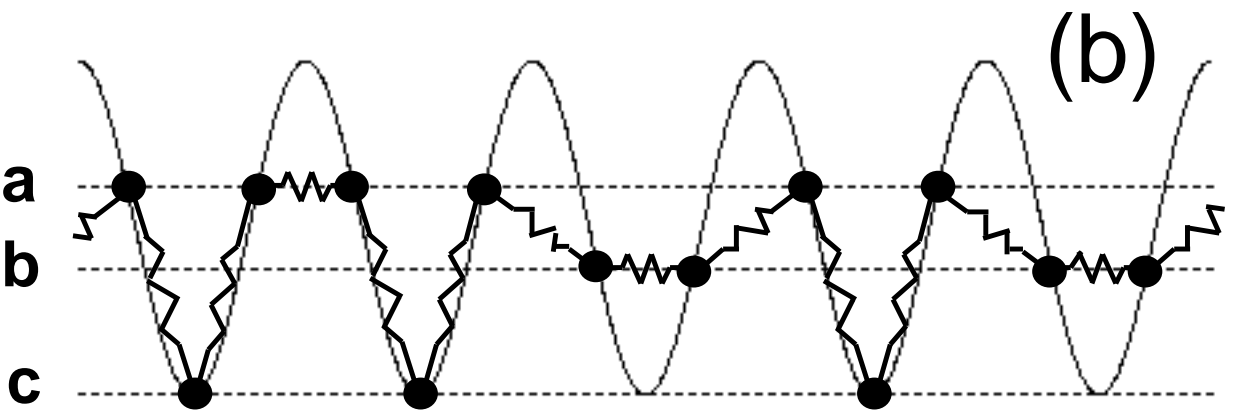
\end{document}